\documentclass[a4paper,fleqn]{cas-dc}

\usepackage[numbers]{natbib}

\def\tsc#1{\csdef{#1}{\textsc{\lowercase{#1}}\xspace}}
\tsc{WGM}
\tsc{QE}

\begin{document}
\let\WriteBookmarks\relax
\def\floatpagepagefraction{1}
\def\textpagefraction{.001}
\let\printorcid\relax 

\shorttitle{A New Ultrafast Printer for Large-Scale Assembly of Piezoelectric Biomaterials}    

\shortauthors{Zhengbao Yang et al.}

\title[mode = title]{A New Ultrafast Printer for Large-Scale Assembly of Piezoelectric Biomaterials}

\author[1]{Nan An}

\author[1]{Mingtong Chen}

\author[1]{Zhengbao Yang}
\cormark[1] 
\ead{zbyang@hk.ust} 
\ead[URL]{https://yanglab.hkust.edu.hk/}

\address[1]{The Hong Kong University of Science and Technology
Hong Kong, SAR 999077, China}

\cortext[1]{Corresponding author} 

\begin{abstract}
We propose a modular, fast and large-area fabrication of bio-piezoelectric films. The technique is based on the formation of cone-jet mode by applying a high voltage electric field to conductive spiked metal disks. And the self-assembly process of biomolecular materials through nanoconfinement with in-situ poling effect. This job achieved print speeds of up to 9.2 × 10\textsuperscript{9} $\mu$m\textsuperscript{3}/s with a combination of only 2 printheads. At the same time, the modular design allows the MLSP to achieve theoretically unlimited print efficiency. It also provides flexible configuration options for different printing needs, such as preparing films of different areas and shapes. In short, MLSP demonstrates the ability of piezoelectric biomaterials to undergo ultra-fast, large-scale assembly. Demonstrates good potential as a universally applicable bio-device for the fabrication of bio-piezoelectric films.

\end{abstract}




\maketitle

\section{Introduction}

Since the discovery of the piezoelectric effect in single-crystal Rochelle salt by the Curie brothers in 1880, it has been extensively and intensively studied\cite{1}. This effect allows for the conversion of electrical and mechanical energy, and is based on the intrinsic properties of crystalline materials with non-neutral symmetry\cite{2}. When subjected to a mechanical force, the piezoelectric material deforms, causing the centers of positive and negative charges to change with the direction of the force, resulting in a spontaneous polarization effect. This causes positive and negative charges to collect on the surface of the piezoelectric material, resulting in a piezoelectric potential\cite{3}. For more than a century, the study of the piezoelectric effect has led to numerous applications in the fields of actuators, sensors, energy harvesting and catalysis\cite{4,5,6,7,8}. Ceramic piezoelectric materials, represented by zirconium titanate (PZT), have been widely commercialized, but the inherent toxicity of lead in the raw material poses serious biological and environmental hazards during production, use and recycling\cite{9}. Although lead-free ceramic piezoelectric materials solve the above problems, the rigidity of inorganic piezoelectric materials limits their use in highly flexible and dexterous scenarios\cite{10}. Natural piezoelectric biomaterials are seen as a solution to the above problems. In addition, the biocompatibility and biodegradability of natural piezoelectric biomaterials give them additional advantages over piezoelectric ceramics and polymers, making them ideal for use in a wide range of medical and biological applications\cite{11}.

The piezoelectric effect is present in a wide range of biological tissues, such as blood vessels, tendons, skin, hair, cornea, bone, wood, etc. \cite{12,13,14,15,16,17}. A variety of biomolecules are also piezoelectric, such as amino acids, peptides, proteins, DNA, cellulose, chitosan, chitin, etc. \cite{11,18,19,20}. However, the inherent complex structure of various biomolecules materials poses a challenge for their piezoelectric fabrication and application. Glycine is the only non-chiral amino acid and the simplest amino acid that can be crystallized into three different polycrystalline species, namely $\alpha$, $\beta$ and $\gamma$-glycine\cite{21}. It has been shown that $\beta$ and $\gamma$ glycine exhibit extremely strong piezoelectricity\cite{22}. As a result, numerous possibilities for the fabrication and application of biomolecular structures have been attempted using glycine as a raw material. Hosseini et al. proposed a Glycine-Chitosan-based $\beta$-glycine crystal film\cite{23}. Yang et al.  developed a sandwich-structured $\gamma$-glycine crystal film using polyvinyl alcohol (PVA) as a substrate\cite{10}. However, existing self-assembly strategies still have many limitations in the fabrication of glycine piezoelectric materials, with domain alignments typically taking a long (24-48 hour) period, and limited control of active crystallization during the crystallization of biomolecules\cite{23,24}. This undoubtedly limits the possibility of large-scale assembly and realization of macroscopic piezoelectricity in biomolecules materials.

Here, we have designed a modularized large-scale super-fast printing strategy (MLSP) that can be used for the fabrication of piezoelectric thin films. It is based on the electrohydrodynamic spray method. The printer's nanoconfinement and in-situ poling effects enable biomolecules to achieve autonomous domain alignment and immobilize on the substrate to achieve a large-area self-assembly process without the need for polarization after the film is made. Modular printhead and ink supply piping enables macro-piezoelectricity and programmable manufacturing of piezoelectric films. And it provides scalability for the fabrication of piezoelectric films for different needs. In real-world testing, the MLSP Printer achieves (i) Speeds that far exceed conventional piezoelectric thin film fabrication techniques, with deposition speeds of up to 9.2×10\textsuperscript{9}$\mu$m\textsuperscript{3}/s. And modules can be stacked as needed to reach theoretical unlimited print rates. (ii) The resulting $\beta$-glycine film has excellent piezoelectric properties (d\textsubscript{33}=8.5 pm/V). MLSP demonstrates the feasibility of large-scale preparation of piezoelectric biomaterials and shows good prospects for them to become truly practical biological devices.

\section{Result}

We designed an MLSP printer consisting of multiple homemade printheads, ink supply pipe fabricated using 3D printing while acting as a support design, pumps for ink supply, a roll-to-roll deposition platform with heating capabilities, and a power supply connected between the printheads and the deposition platform (Figure 1a). The printhead consists of two parts, a round metal disk with spikes and a dispensing needle with a 90° bend in the tip. The circular metal disk is surrounded by 16 spikes, and the diameter of the spiked metal disk is 10mm (dedendum circle), with a thickness of 20$\mu$m. The center is hollowed out and secured to the tip of the dispensing needle by a conductive metal adhesive. Selecting a wider diameter dispensing needle (18G) can effectively avoid clogging and can also support inks with higher viscosity. The ink supply pipe manufactured using 3D printing is divided into three parts. The head pipe is connected to the ink supply pump and also has a fixing function for securing the entire print section to the printer. Diverter piping that directs ink to the printheads. The number of manifolds can be increased or decreased, and the design can be changed according to actual needs. The luer lock is responsible for securing the printheads. The unthreaded tapered end glued to the manifold using high-strength waterproof AB adhesive. The female threaded end connected to the dispensing needle, easy to replace while achieving good water resistance. Realization of modular design of printheads and ink supply ducts. The Roll-to-Roll platform utilizes a flexible conductive foil substrate as a self-assembly platform for piezoelectric films. It is also equipped with a heating function to further increase the printing speed of the films.

Figure 1a shows the complete structure of the NLSFP printer, after assembling the NLSFP printer, the printheads are connected by wires for applying a high voltage electric field. Since the metal syringe portion of the dispensing nozzle is closer to the deposition platform than the spiked metal disk of the printhead, the metal syringe below the spiked metal disk is insulated. To avoid the electrostatic field of the metal syringe affecting the printing, and to avoid the high voltage electric field breaking through the air causing the high voltage power supply to go into overload protection.

MLSP printers use electrohydrodynamic spraying to realize the printing of thin films. By applying an electric field between the print nozzle and the deposition platform, a cone-jet mode is formed on the spines of the spiked metal disk, and Figure 2a shows the actual appearance of the cone-jet mode through an optical photograph. 

\begin{figure*}[h]
	\centering
		\includegraphics[scale=1]{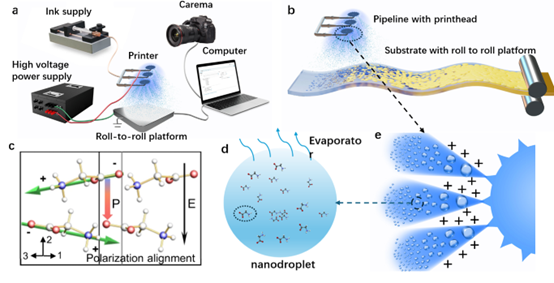}
	  \caption{Structure and Working Principle of the MLSP Printer:a. The MLSP printer consists of the main printing body equipped with a roll-to-roll deposition platform, ink supply system, and high-voltage power supply. It is also integrated with a high-resolution camera for monitoring the printing process and a control computer. b. Schematic illustration of the MLSP printer mechanism for piezoelectric film fabrication. c. Polarization direction of $\beta$-glycine crystals. d.Crystallization process of glycine based on nanoconfinement. e. Schematic diagram of the electrohydrodynamic spraying process.}
      \label{FIG:1}
\end{figure*}

\begin{figure*}[h]
	\centering
		\includegraphics[scale=1]{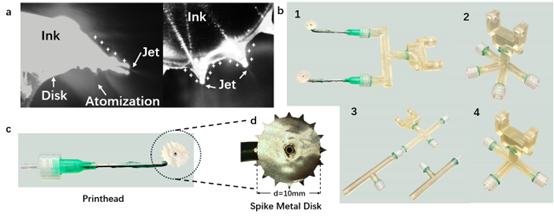}
	  \caption{Printing Results and Modular Components
a. Optical image of the electrohydrodynamic spraying process captured under a microscopic camera. b. Various modular component designs developed for different printing needs and scenarios: 1. Parallel pipeline design. 2. 60° angled pipeline design. 3. Modular and reconfigurable pipeline design. 4. 90° angled pipeline design. c. Optical image of the printhead. d. Optical image of the spiked metal disk captured under a microscopic camera.
}\label{FIG:2}
\end{figure*}

The Cone-jet mode consists of two phases, the first phase is the conical meniscus style of fluid that gathers at the tip of the spike, known as the Taylor cone, formed by the interaction between the electrostatic force at the tip of the spike and the surface tension of the fluid. When the liquid passes through the Taylor cone, the electrostatic charge accumulates at the tip of the spike, forming a tangential electric stress that accelerates the flow of the liquid toward the tip of the spike and forms the second stage of the cone-jet mode, the steady jet ejection. As the jet moves forward, the liquid becomes jet unstable driven by tangential electrical stresses, causing the liquid column of the jet to break up and release parent droplets proportional to the diameter of the jet. The parent droplet continues to spread under the drive and inertia of the electric field. During diffusion, the liquid in the parent droplet continues to evaporate, causing the charge held on the surface of the droplet to reach the Rayleigh limit. For a spherical droplet, the expression for the Rayleigh limit is:

$$
q=\sqrt{8\pi^2\varepsilon_0\gamma d^3} 
$$

Under the interaction of Coulomb force and surface tension, the parent droplet that reaches the Rayleigh limit undergoes Coulomb fission to form smaller sub-droplets. The process is continuously recycled to eventually form nanoscale droplets and then aerosolized.

Having a bio-ink with stable dispersibility is essential to guarantee the realization of electrohydrodynamic sprays. Inks for printing biofilms were prepared in the ratio of 10:1:2.5 = deionized water: glycine: ethanol. Ethanol is added to increase the ionic content of the ink to increase the conductivity of the ink, while the rapid evaporation of ethanol during the printing process also helps to increase the printing speed of the film. In order to achieve a stable printing effect, the electrostatic field voltage between the printhead and the deposition platform was set at 7kV-10kV, and the average distance between the barbed metal disk of the printhead and the deposition platform was 20mm. Having the above conditions, the printer achieved the formation of a stable electrohydrodynamic spray and started the preparation of glycine piezoelectric films. During spraying of glycine inks, as the solvent evaporates, droplets rupture and the ink forms a supersaturated solution, leading to a nanoconfinement effect that constrains glycine to form $\beta$-phase polycrystalline species for nucleation. During nucleation, the high-voltage electric field applied between the nozzle and the deposition platform guides the domain alignment of the crystals in the spray, culminating in the formation of a net polarization direction parallel to the electric field [0, 2, 0], a process known as in-situ poling. The $\beta$-glycine nanocrystals were deposited onto the Roll-to-Roll platform, but at this time the periphery of the crystals was still wrapped by the incompletely evaporated solution. Through the heating of the platform the nanocrystals were rapidly aggregated and with the evaporation of the solution to form dense $\beta$-glycine nanocrystal piezoelectric films. 

To enable large-scale and modular printing, we developed various modular configurations for the MLSP printer (Figure 2b), including parallel dual/multi-printhead designs, multi-printhead arrangements at 60° and 90° angles, and a modular pipeline system allowing flexible addition or removal of printheads. These diverse modular pipeline systems enable adjustments to the printing area and printing speed, which are critical for achieving high-efficiency printing. However, multi-printhead configurations may introduce challenges in printing rate and quality. Therefore, identifying an optimal spatial arrangement of the printheads is essential. Piezoelectric properties of bio-piezoelectric films.
In electrohydrodynamic spraying, achieving a stable cone-jet mode requires a continuous and steady current. The magnitude of the current (I) determines the volumetric flow rate (Q), especially for liquids with high electrical conductivity (K).

$$
I=f(\varepsilon_r)\sqrt{\frac{\gamma QK}{\varepsilon_r}} 
$$

The process in which the liquid accumulates at the tip and is sprayed under the influence of tangential electric stress in the cone-jet mode is referred to as the tip effect. To investigate the mutual influence of electric fields between printheads, identify an optimal printhead arrangement, and better visualize how the tip effect governs liquid spraying, electrostatic field simulations were conducted.

\begin{figure*}[h]
	\centering
		\includegraphics[scale=1]{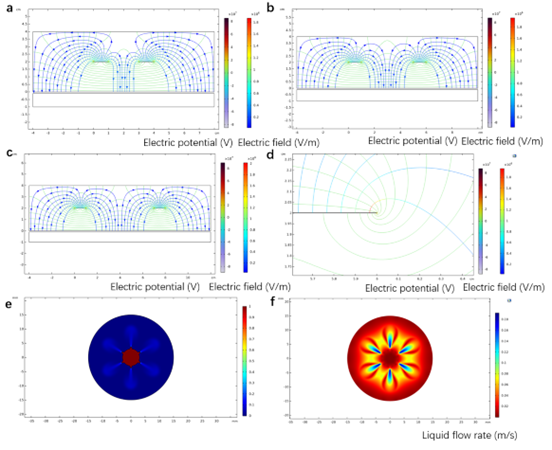}
	  \caption{Electrostatic field and electrohydrodynamic simulations:a. Cross-sectional electrostatic field simulation result at a printhead spacing of 3mm under an applied voltage of 8kV. b. Cross-sectional electrostatic field simulation result at a printhead spacing of 5mm under an applied voltage of 8kV. c. Cross-sectional electrostatic field simulation result at a printhead spacing of 7mm under an applied voltage of 8kV. d. Distribution of electric potential and electric field at the printhead tip. e. Simulated volume fraction in the electrohydrodynamic simulation of a simplified printhead. f. Simulated fluid ejection velocity in the electrohydrodynamic simulation of a simplified printhead.}
      \label{FIG:3}
\end{figure*}

\begin{figure*}[h]
	\centering
		\includegraphics[scale=1]{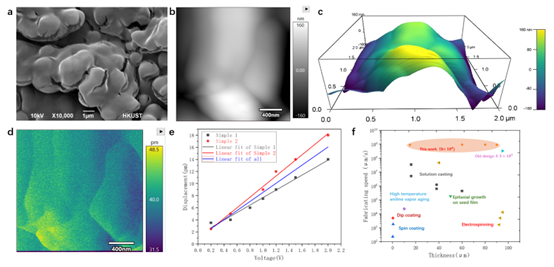}
	  \caption{Structural characterization, PFM measurements, and printing efficiency:a. Surface morphology SEM image showing densely packed crystal domains forming a uniform and continuous film. b. Two-dimensional PFM topography of the film. c. Three-dimensional PFM topography of the $\beta$-glycine crystal film. d. PFM amplitude map of the glycine film. e. Linear dependence of PFM amplitude on the applied AC voltage; the slope of the line provides the effective piezoelectric coefficient of the $\beta$-glycine crystal film. f. Rate comparison of different printing strategies.}
      \label{FIG:4}
\end{figure*}

Figures 3a, 3b, and 3c clearly illustrate the mutual influence of electric fields between printheads at different spacing intervals. When the spacing between two printheads is 3cm, the high-voltage electric field significantly disrupts the tip effect, resulting in a weakened electric field directed from the printhead to the grounded substrate. As demonstrated experimentally, the affected tip fails to form a cone-jet mode, and the liquid cannot be sprayed. When the spacing is adjusted to 7cm, as shown in Fig. 3c, the mutual interference between printheads is nearly eliminated, allowing the tip effect to function properly and enabling normal electrohydrodynamic spraying.

Figure 3d specifically shows the electric field intensity and current characteristics under the tip effect. Figures 3e and 3f present electrohydrodynamic simulations of a simplified single-printhead model equipped with a spiked metal disk containing six spikes, visually demonstrating the spraying behavior driven by the tip effect.

The structure of the fabricated glycine piezoelectric films was investigated using a Scanning Electron Microscope (SEM). The microscopic surface morphology of the film is shown in Fig. 4a. Crystalline glycine domains can be observed densely packed together, forming a continuous film structure.

The piezoelectric properties of the fabricated $\beta$-glycine crystal films were analyzed using piezoresponse force microscopy (PFM). As shown in Fig. 4b, the surface morphology of the film was obtained via PFM. The uniform contrast in both the phase and amplitude images indicates a well-defined crystalline domain structure. Figure 4c presents the 3D surface morphology of the $\beta$-glycine crystal in the selected region.
By applying different AC voltages to various regions of the sample, the out-of-plane (OOP) amplitude response was recorded, as shown in the figure. The OOP amplitude increases linearly with the applied AC voltage. By fitting the piezoelectric data from two different regions of the glycine film, an average effective piezoelectric coefficient of approximately 7.5pm/V was obtained, which is higher than the theoretical value (5.7pm/V), demonstrating the excellent piezoelectric performance of the $\beta$-glycine crystal films fabricated using the MLSP method.
In addition, we compared the volume of bio-piezoelectric materials fabricated using existing common technologies as a function of film thickness with the glycine films produced in this work. The range of printing speeds is shown in Figure 4f. The figure presents a plot of deposition rate versus achievable film thickness across different fabrication techniques for piezoelectric films. The data in the figure includes two categories of deposition technologies: Vapor-phase deposition, including pulsed laser deposition and magnetron sputtering; and solution-chemistry-derived deposition, including sol-gel, composite sol-gel, aerosol deposition, and electrophoretic deposition, as well as traditional electrohydrodynamic (EHD) spraying. As a general trend, solution-chemistry-derived methods are capable of producing thicker films at faster rates compared to vapor-phase methods. Under this comparison, the maximum printing speed of the MLSP printer reaches up to 9.2 × 10\textsuperscript{9} $\mu$m\textsuperscript{3}/s.

\section{Method}

The composition of the MLSP includes self-made multiple printheads. Using laser cut stainless steel spiked discs secured by conductive metal adhesive to the dispensing needle (18G) with a 90° bend in the syringe. The metal syringe and metal conductive adhesive are insulated except for the part connected to the wire. Modular design of the ink supply piping, divided into header piping, diverter piping using light-curing 3D printing and luer lock. The three are combined together using a high-strength waterproof structural AB adhesive. The ink supply pipe is connected to the printhead through the threads of the luer lock. The ink supply pump is connected to the head pipe and is used to supply ink to the printheads and control the flow rate. The high voltage power supply connects the printheads via wires to provide a positive charge to the printer. Meanwhile, the Roll-to-Roll platform's flexible conductive foil substrate is grounded. The speed of substrate movement can be controlled in real time according to printing needs for uniform, large-area film manufacturing.

2g of glycine powder was dissolved in 20 ml of deionized water and 5 ml of ethanol was added to increase ink conductivity. The mixture was stirred at room temperature (25℃) for 10 minutes using a magnetic stirrer to obtain a homogeneous solution. The obtained glycine solution was placed directly into the ink supply pump of the MLSP printer for the printing of piezoelectric films. During electrohydrodynamic atomization, a strong electric field on the printhead is applied to the stinger tips of the barbed metal disks, forming a cone-jet mode. Bursting into tiny droplets driven by tangential electrical stresses. Eventually nucleation of $\beta$-glycine crystals in the presence of nanoconfinement. For the preparation of glycine films, the distance between the metal disk of the printhead and the substrate was 20 mm, the applied voltage was 7-10 kV.










\bibliographystyle{cas-model2-names}

\bibliography{cas-refs}



\end{document}